\documentclass[review]{elsarticle}

\usepackage{lineno,hyperref}
\usepackage{color}
\modulolinenumbers[5]

\journal{Chemical Physics Letters}

\begin{document}

\begin{frontmatter}

\title{Electrical transport and magnetic properties of the triangular-lattice compound Zr$_2$NiP$_2$}

\author{Zongtang Wan,  Yuqian Zhao, Jiabin Liu}

\author{Yuesheng Li\corref{mycorrespondingauthor}}
\cortext[mycorrespondingauthor]{Corresponding author}
\ead{yuesheng\_li@hust.edu.cn}

\address{Wuhan National High Magnetic Field Center and School of Physics, Huazhong University of Science and Technology, 430074 Wuhan, China}

\begin{abstract}
We report the first investigation of the electrical and magnetic properties of the triangular-lattice compound Zr$_2$NiP$_2$ (space group $P$6$_3$/$mmc$). The temperature evolution of electrical resistivity follows the Bloch-Gr\"uneisen-Mott law, and exhibits a typically metallic behavior. No transition is visible by both electrical and magnetic property measurements, and nearly no magnetization is detected ($M_0$ $<$ 0.002$\mu_\mathrm{B}$/Ni) down to 1.8 K up to 7 T. The metallic and nonmagnetic characters are well understood by the first-principles calculations for Zr$_2$NiP$_2$. 
\end{abstract}


\end{frontmatter}

\section{Introduction}

Quasi-two-dimensional materials with exotic properties have fascinated condensed matter physicists for decades. Among others, the prototype of a resonating-valence-bond quantum spin liquid state had been proposed by P. W. Anderson on the $S$ = 1/2 triangular-lattice Heisenberg antiferromagnetic model in 1973~\cite{ANDERSON1973153}. The experimental realizations of $S$ = 1/2 triangular-lattice spin-liquid states had been extensively reported, e.g., in $\kappa$-(BEDT-TTF)$_2$Cu$_2$(CN)$_3$~\cite{PhysRevLett.91.107001,yamashita2008thermodynamic}, EtMe$_3$Sb[Pd(dmit)$_2$]$_2$~\cite{PhysRevB.77.104413}, Ba$_3$CuSb$_2$O$_9$~\cite{PhysRevLett.106.147204},  YbMgGaO$_4$~\cite{li2015gapless,li2015rare}, and  NaYbCh$_2$ (Ch = O, S, Se)~\cite{weiwei2018Rare,PhysRevB.98.220409}. Recently, $S$ = 1 triangular-lattice spin-liquid ground states were also proposed~\cite{PhysRevB.81.224417,PhysRevB.84.180403,PhysRevLett.108.087204,PhysRevB.86.224409}, and might be materialized by the Ni-based compounds, including NiGa$_2$S$_4$~\cite{Nakatsuji2005Spin,Nakatsuji2010Novel} and Ba$_3$NiSb$_2$O$_9$~\cite{PhysRevLett.107.197204}. Besides this, unconventional high-temperature superconductors had been discovered in many quasi-two-dimensional compounds, including Cu-based~\cite{Bednorz1986Possible,PhysRevLett.58.908} and Fe-based~\cite{Kamihara2008Iron} systems. It has been proposed for a long time that the high-temperature superconductivity can be considered as evolving from a doped spin-liquid state~\cite{anderson1987resonating,RevModPhys.78.17}. The ``smoking gun'' evidence for the intimate relationship between the high-temperature superconductivity and spin liquid remain absent, and most of the existing studies on spin liquid are restricted in the quasi-two-dimensional geometrically frustrated Mott insulators. Therefore, it is intriguing to explore the physical properties of new Cu-, Ni- or Fe-based metals with quasi-two-dimensional geometrically frustrated lattices, e.g., the triangular lattice.

The Ni-based triangular-lattice compound Zr$_2$NiP$_2$ belongs to the family of the ternary pnictides Ln$_2$NiX$_2$  (Ln = Tb, Dy, Ho, Er and Zr; X = P, As), with the hexagonal modification ($P$6$_3$/$mmc$ space group)~\cite{Ghadraoui1988new}.  Although these compounds have been synthesized and structurally characterized long time ago~\cite{Ghadraoui1988new}, other physical properties, e.g., electrical conductivity and magnetism, remain completely unknown. In this work, we conducted the electrical transport and magnetic property characterizations of Zr$_2$NiP$_2$. As shown in Figure~\ref{fig1}, Ni atoms form a geometrically perfect triangular lattice (intralayer Ni-Ni distance of $\sim$ 3.77 \AA), and are spatially separated by the double layers of ZrP$_6$ octahedra (interlayer Ni-Ni distance of $\sim$ 6.44 \AA). The temperature dependence of electrical resistivity is reasonably described by the Bloch-Gr\"uneisen-Mott (BGM) law, and indicates the intrinsic metallic property of Zr$_2$NiP$_2$. Neither structural nor magnetic transition is observed by electrical resistivity and magnetic susceptibility down to 1.8 K. The low-$T$ magnetization is measured to be tiny, $M_0$ $\le$ 0.002$\mu_\mathrm{B}$/Ni, up to 7 T. Ni should be in the nonmagnetic state, which accounts for the weak magnetism observed in Zr$_2$NiP$_2$. 

\section{Materials and methods}

\begin{figure}[t]
\begin{center}
\includegraphics[width=15cm,angle=0]{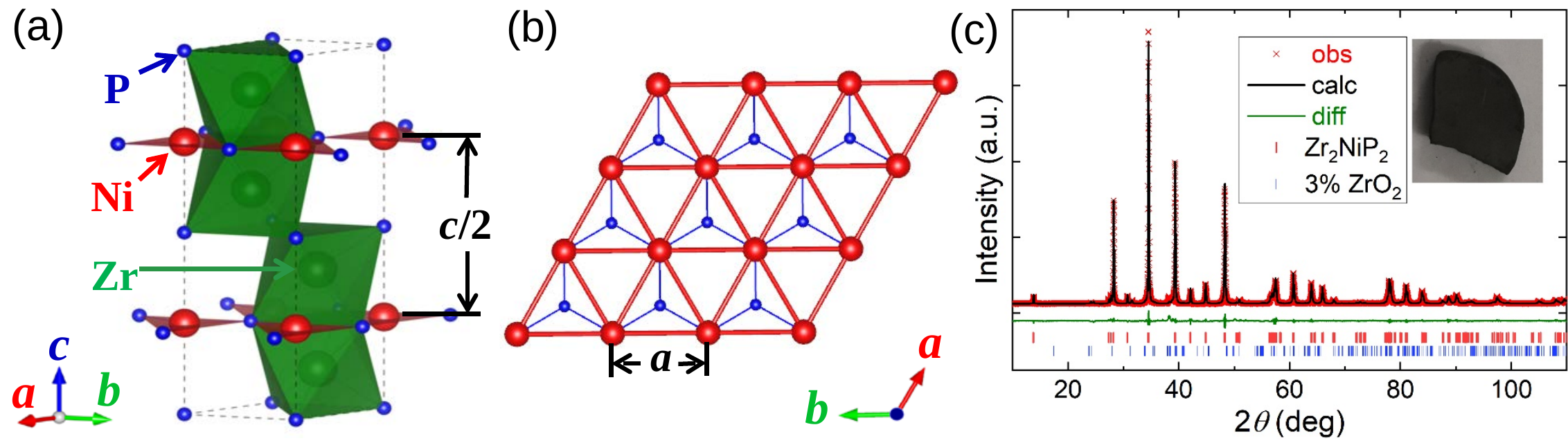}
\caption{
(a) Crystal structure of Zr$_2$NiP$_2$. NiP$_3$ triangles (red) and ZrP$_6$ octahedra (olive) are shown, and the dashed black lines mark the unit cell. (b) The regular triangular lattice of Ni atoms. The interlayer distance of Ni-Ni is $c$/2 $\sim$ 6.443 \AA, much larger than the intralayer one $a$ $\sim$ 3.772 \AA. (c) X-ray powder diffraction measured at 300 K and Rietveld refinement for Zr$_2$NiP$_2$. The inset shows a typical as-grown sample of Zr$_2$NiP$_2$.}
\label{fig1}
\end{center}
\end{figure}

The polycrystalline samples of Zr$_2$NiP$_2$ were synthesized from stoichiometric mixtures of zirconium (99.5\%, aladdin), nickel (99.8\%,alfa), and red phosphorus (99.999\%, aladdin). To prevent oxidation, the starting materials were mixed and pressed into pellets in an argon atmosphere glove box. The pellets wrapped in platinum foil were transferred into a quartz tube,  which were sealed under vacuum (inner pressure of $\sim$ 2$\times$10$^{-9}$ bar). The sealed tube was heated to 500 $^\circ$C at a rate of 1 $^\circ$C/min, and its temperature was maintained for three days. After that, the tube was further heated to 1000 $^\circ$C for seven days, and then cooled down to room temperature at a rate of $\sim$ $-$1 $^\circ$C/min. Air-insensitive and polycrystalline black Zr$_2$NiP$_2$ samples were obtained, as shown in Figure~\ref{fig1}(c).

The phase purity of the sample was checked by powder x-ray diffraction (XRD, Rigaku) measurements with Cu $K_\alpha$ radiation ($\lambda$ = 1.5418 \AA), and a small weight fraction of nonmagnetic impurity ZrO$_2$, 3.2$\pm$0.1\%, was found from Rietveld refinement (see Figure~\ref{fig1}(c)), possibly due to the short exposure to air during the synthesis process. The reaction between the mixtures and the quartz tube was prevented by the platinum foil. The General Structure Analysis System (GSAS) program was used for the Rietveld refinements~\cite{gsas}. The ratio of Zr:Ni:P was semi-quantitatively determined by a scanning electron microscope equipped with a x-ray energy dispersive spectrometer (SEM–EDX, JEOL) to be $\sim$ (2.0$\pm$0.3):1:(1.9$\pm$0.3), consistent with the ideal ratio of 2:1:2 within the resolution. The electrical resistivity down to 2 K was measured by using the standard four-probe ac technique on two cuboid-shaped as-grown samples of Zr$_2$NiP$_2$ (S$_1$ and S$_2$) in a physical property measurement system (PPMS, Quantum Design) at both 0 and 9 T. The dc magnetization measurements were carried out in a magnetic property measurement system (MPMS, Quantum Design) down to 1.8 K up to 7 T, by using the as-grown samples of $\sim$ 40 mg. As the magnetization signal of the title compound is very weak, we carefully chose a quartz holder without evident background in the magnetic property measurements. 

We conducted first-principles calculations for Zr$_2$NiP$_2$ based on the density functional theory (DFT) with the generalized gradient approximation in the Vienna Ab initio Simulation Package (VASP)~\cite{vasp1,vasp2,PhysRevLett.77.3865}. The crystal structure experimentally determined by XRD was used as the starting one, and the kinetic energy cutoff for the plane waves was specified to be 330 eV.  The 6$\times$6$\times$6 $k$-mesh was used, and the lattice parameters were fixed to their experimental values. The residual forces were below 0.003 eV/\AA~and the total energy was converged to 10$^{-8}$ eV in the fully optimized structure with $z$(Zr) = 0.10995. We performed both DFT simulations with Coulomb repulsions $U$ = 0 and 5 eV on Ni sites, but no essential difference was found. Furthermore, we also relaxed the lattice parameters simultaneously, and obtained $a$ = 3.775 \AA, $c$ =  13.07 \AA, and $z$(Zr) = 0.10996, reasonably consistent with the experimental result (see Table~\ref{tab1}).   

\section{Results and discussion}

\subsection{Crystal structure}

\begin{table}
\caption{Crystal structure of the Zr$_2$NiP$_2$ phase determined by XRD at 300 K. The space group is $P$6$_3$/$mmc$ (No.194) with $a$ = 3.7717(1) {\AA} and  $c$ = 12.8860(4) {\AA}, and the final Rietveld refinement results in $R_\mathrm{p}$ = 2.46\% and $R_\mathrm{wp}$ = 3.99\%. The weight fraction of the impurity phase ZrO$_2$ (space group $P$ 2$_1$/$c$) is refined to be, 3.2$\pm$0.1\%. Due to the low intensities of the reflections of the impurity phase (Figure~\ref{fig1}(c)), we only refined the phase fraction and lattice parameters  of ZrO$_2$, but fixed other atom parameters as reported in reference~\cite{smith1965crystal}. The resulted $a$(ZrO$_2$) = 5.143(2) {\AA}, $b$(ZrO$_2$) = 5.285(3) {\AA}, $c$(ZrO$_2$) = 5.373(3) {\AA}, and $\beta$(ZrO$_2$) = 99.17(2)$^\circ$ are reasonably consistent with the reported result of ZrO$_2$~\cite{smith1965crystal}.}
\begin{center}
\begin{tabular}{ c c c c c c c}
    \hline
    \hline
 Atoms & Wyckoff position & $x$ & $y$ & $z$ & Occ. & 100$\times U_{\mathrm{iso}}$ (\AA$^2$)\\ \hline
Zr & 4f & 1/3 & 2/3 & 0.10995(4) & 1 & 1.10(2) \\
Ni & 2b	& 0 & 0 & 1/4 & 1 & 2.34(5) \\
P1 & 2a & 0 & 0 & 0 & 1 & 1.22(7) \\
P2 & 2c & 1/3 & 2/3 & 3/4 &1 & 1.75(6) \\
    \hline
    \hline
\end{tabular}
\end{center}
\label{tab1}
\end{table}

Figure~\ref{fig1}(c) shows the powder XRD pattern measured at 300 K and the final Rietveld refinement. Except the possible nonmagnetic ZrO$_2$ (weight fraction $\sim$ 3\%), no other impurity phases are visible in the samples of Zr$_2$NiP$_2$. The grain size is estimated to be larger than 100 nm, according to the narrow full width at half maximum $\sim$ 0.09$^\circ$~\footnote{The instrumental resolution is about 0.05$^\circ$.}  observed at 2$\theta$ = 34.5$^\circ$ (see Figure~\ref{fig1}(c))~\cite{gsas}. The detailed crystallographic parameters and atomic positions are listed in Table~\ref{tab1}. The refinement result suggests that the title compound has a hexagonal structure and the $P$6$_3$/$mmc$ space group with two formulae per unit cell, as previously reported in reference~\cite{Ghadraoui1988new}. Moreover, our refined crystal structure shows quantitative agreement with the previous result~\cite{Ghadraoui1988new}. Ni atoms occupy the 2b Wyckoff position and form a geometrically perfect triangular lattice (see Figure~\ref{fig1}(b)). Along the $c$ axis, the triangular layers of Ni atoms are stacked in an ``AA''-type stacking fashion, and well separated by  the double layers of ZrP$_6$ octahedra. The interlayer Ni-Ni  distance is $c$/2 = 6.4430(2) \AA, much larger than the intralayer Ni-Ni distance of $a$ = 3.7717(1) \AA. The Ni-Zr distances $\sim$ 2.8  \AA~are reasonably short, which may have considerable influence on the magnetism of Zr$_2$NiP$_2$. All of these indicate a quasi-two-dimensional nature of the system of Zr$_2$NiP$_2$. Moreover, no antisite mixing was reported~\cite{Ghadraoui1988new} due to the large chemical difference among Zr, Ni, and P atoms, and thus the structural disorder should have marginal effects in Zr$_2$NiP$_2$. 

\subsection{Electrical transport} 

\begin{figure}[t]
\begin{center}
\includegraphics[width=15cm,angle=0]{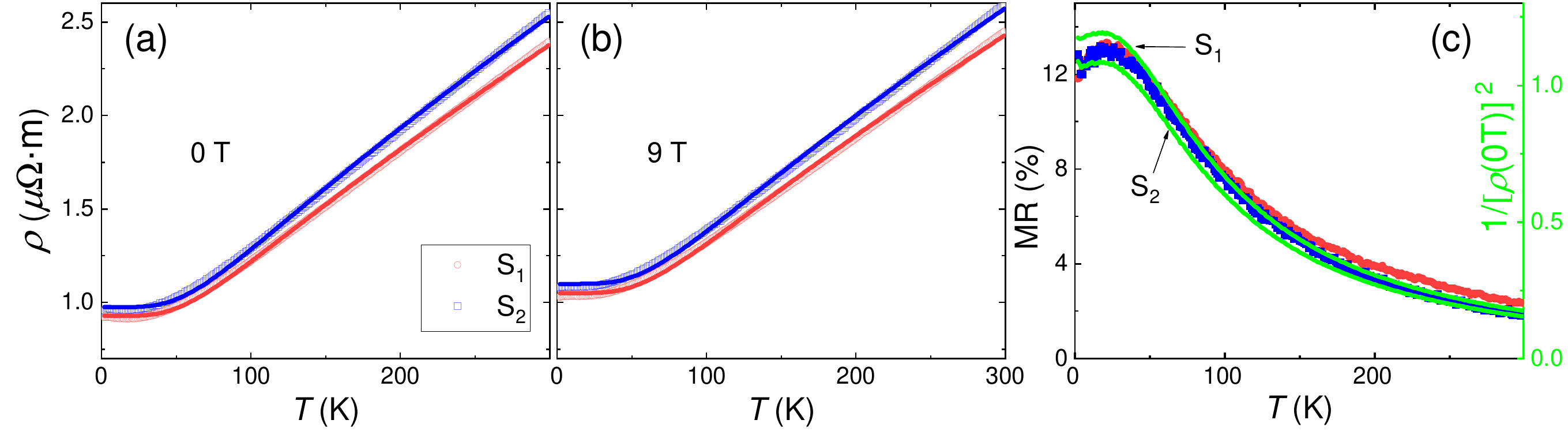}
\caption{
Temperature dependence of the electrical resistivity ($\rho$) measured at (a) 0 and (b) 9 T on two as-grown samples of Zr$_2$NiP$_2$ (S$_1$ and S$_2$). The colored lines present the BGM fits (see main text) from 2 to 300 K. (c) The magnetoresistance of Zr$_2$NiP$_2$ measured at 9 T, MR(9 T) = [$\rho$(9 T)$-\rho$(0 T)]/$\rho$(0 T). The green lines show 1/[$\rho$(0 T)]$^2$ measured on S$_1$ and S$_2$, respectively.}
\label{fig2}
\end{center}
\end{figure}

\begin{table}
\caption{The BGM fitting parameters for  Zr$_2$NiP$_2$. Here, $\sigma_\mathrm{LS}$ presents the least-squares standard deviation.}
\begin{center}
\begin{tabular}{ c c c c c c }
    \hline
    \hline
  & $\rho_0$ ($\mu\Omega\cdot$m) & $\Theta_\mathrm{D}$ (K) & $R$ ($\mu\Omega\cdot$m/K) & $K$ ($\mu\Omega\cdot$m/K$^3$) & $\sigma_\mathrm{LS}$ ($\mu\Omega\cdot$m) \\\hline
S$_1$, 0 T &  0.93 & 342 & 0.0052 & 4$\times$10$^{-10}$ & 0.0088 \\
S$_2$, 0 T &  0.98 & 349 & 0.0057 & 1$\times$10$^{-9}$ & 0.0108 \\\hline
S$_1$, 9 T &  1.05 & 363 & 0.0051 & 8$\times$10$^{-10}$ & 0.0109 \\
S$_2$, 9 T &  1.10 & 364 & 0.0054 & 8$\times$10$^{-10}$ & 0.0122 \\
    \hline
    \hline
\end{tabular}
\end{center}
\label{tab2}
\end{table}

Figure~\ref{fig2}(a) and (b) show the electrical resistivity ($\rho$) measured on two independent as-grown samples of Zr$_2$NiP$_2$ at $\mu_0H$ = 0 and 9 T  down to 2 K, and no significant sample dependence is observed. $\rho$ decreases with the decrease of temperature, and gradually reaches a minimum below $\sim$ 25 K. The measured temperature dependence can be well described by the BGM law~\cite{mott1958the,Grimval1981the}, 
\begin{equation}
\rho(T)=\rho_0+4R\Theta_\mathrm{D}(\frac{T}{\Theta_\mathrm{D}})^5\int_0^{\Theta_\mathrm{D}/T}\frac{x^5dx}{(e^x-1)(1-e^{-x})}-KT^3,
\label{eq1}
\end{equation}
where $\rho_0$ is the temperature-independent residual resistivity, the second term accounts for the electron-phonon scattering, and the $T^3$ term is due to the Mott-type $s$-$d$ electron scattering. All the least-squares fitting parameters are listed in Table~\ref{tab2}. The resulted $\rho_0$, Debye temperature $\Theta_\mathrm{D}$, and electron-phonon coefficient $R$ of Zr$_2$NiP$_2$ show typically metallic features~\cite{PhysRevB.54.9891,Pikul2003}.  In comparison, the coefficient $K$ that is directly related to the strength of the $s$-$d$ interaction is about one order of magnitude smaller than those of many other metals~\cite{PhysRevB.54.9891,Pikul2003}.

Figure~\ref{fig2}(c) shows the magnetoresistance measured on Zr$_2$NiP$_2$ at 9 T, MR(9 T) = [$\rho$(9 T)$-\rho$(0 T)]/$\rho$(0 T), which also exhibits the common positive values of metals, MR $\sim$ several percent~\cite{Nickel1995Hewlett,Magnetoresistance}.  As presented in Figure~\ref{fig2}(c), the temperature evolution of MR(9 T) roughly follows that of 1/[$\rho$(0 T)]$^2$ in the case of Zr$_2$NiP$_2$. To understand this observation, we consider both holes and electrons ($n_1$, $q_1$, $\mu_1$, and $n_2$, $q_2$, $\mu_2$, respectively),
\begin{equation}
\mathbf{j}=(\frac{n_1q_1\mu_1}{1+B^2\mu_1^2}+\frac{n_2q_2\mu_2}{1+B^2\mu_2^2})\mathbf{E}+(\frac{n_1q_1\mu_1^2}{1+B^2\mu_1^2}+\frac{n_2q_2\mu_2^2}{1+B^2\mu_2^2})\mathbf{E}\times\mathbf{B},
\label{eq2}
\end{equation}
where the electric field $\mathbf{E}$ and magnetic field  $\mathbf{B}$ are applied perpendicular and parallel to the $z$ axis, respectively. For simplicity,  we assume the carrier density $n_1$ = $n_2$ = $n$, zero-field carrier mobility $\mu_1$ = $-\mu_2$ = $\mu$, and charge  $q_1$ = $-q_2$ = $e$, so that the Hall term is negligible and current density $\mathbf{j}$ is parallel to $\mathbf{E}$. Thereby, the magnetoresistance is given as, MR($B$) =  $B^2\mu^2$ = $B^2$/[2$ne\rho$(0 T)]$^2$, in agreement with the experimental observation.

The electrical transport characterizations (Figure~\ref{fig2}) consistently suggest that Zr$_2$NiP$_2$ is a typical metal without any evident transition down to 2 K. Furthermore, both $\rho$ and low MR imply a nonmagnetic state of Ni in Zr$_2$NiP$_2$. To confirm this, we conducted the magnetic property measurements (see below). 

\subsection{Magnetic properties}

\begin{figure}[t]
\begin{center}
\includegraphics[width=12cm,angle=0]{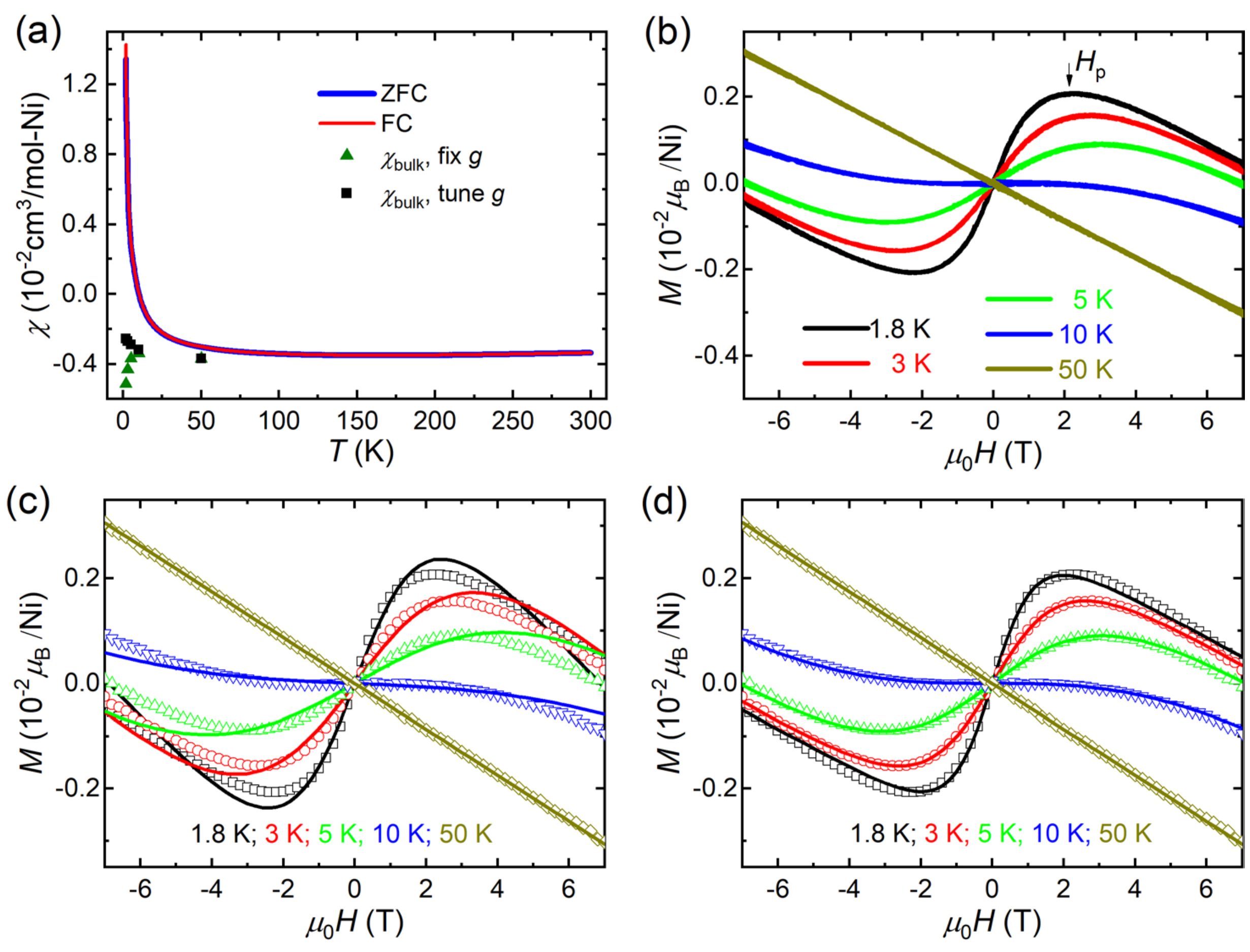}
\caption{
(a) Magnetic susceptibilities measured under zero-field cooling (ZFC) and field cooling (FC) at 0.1 T down to 1.8 K for Zr$_2$NiP$_2$. The scatters show the contributions of the bulk diamagnetism $\chi_\mathrm{bulk}$. (b) Field dependence of isothermal magnetization measured at various temperatures. $H_\mathrm{p}$ at 1.8 K is marked. The Brillouin fits (lines) to the experimental magnetization (scatters) by fixing (c) and tuning (d) the $g$ factor. }
\label{fig3}
\end{center}
\end{figure}

Neither clear anomaly/kink nor splitting between the ZFC and FC susceptibilities is observed (see Figure~\ref{fig3}(a)), which also precludes the existence of magnetic transition in Zr$_2$NiP$_2$ down to 1.8 K. Above $\sim$ 10 K, the magnetic susceptibility is negative, $\chi$ $<$ 0, thus suggesting a diamagnetism. In contrast,  the susceptibility exhibits a profound Curie-like tail at low temperatures, which is most likely to originate from a paramagnetic impurity. To confirm this, we fit the experimental magnetization ($M$) by a combination of a Brillouin function for free $s$ = 1 Ni spins (impurity) and a linear magnetization for the bulk system,
\begin{equation}
M(H,T)=fg\mu_\mathrm{B}\frac{\exp(\frac{2\mu_0Hg\mu_\mathrm{B}}{k_\mathrm{B}T})-1}{\exp(\frac{2\mu_0Hg\mu_\mathrm{B}}{k_\mathrm{B}T})+\exp(\frac{\mu_0Hg\mu_\mathrm{B}}{k_\mathrm{B}T})+1}+\chi_\mathrm{bulk}(T)H,
\label{eq3}
\end{equation}
where $f$ is the fraction of free spins. When the $g$ factor is fixed to the expected value $g$ = 2 with negligible spin-orbit coupling, we obtain $f$ = 0.25\% and $\chi_\mathrm{bulk}$ = $-$0.0052, $-$0.0043, $-$0.0037, $-$0.0034, $-$0.0037 cm$^3$/mol-Ni at 1.8, 3, 5 10, 50 K, respectively, with the least-squares standard deviation $\sigma_\mathrm{LS}$ = 0.0002$\mu_\mathrm{B}$/Ni (see Figure~\ref{fig3}(c)).  Moreover, we also treat $g$ as an adjustable parameter, the goodness of fit gets better ($\sigma_\mathrm{LS}$ = 0.00004$\mu_\mathrm{B}$/Ni), and we find $g$ = 3.3, $f$ = 0.09\%, and $\chi_\mathrm{bulk}$ = $-$0.0026, $-$0.0027, $-$0.0029, $-$0.0032, $-$0.0037 cm$^3$/mol-Ni at 1.8, 3, 5 10, 50 K, respectively (see Figure~\ref{fig3}(d)). Both combined fits suggest that the bulk system of Zr$_2$NiP$_2$ remains diamagnetic/nonmagnetic down to 1.8 K, i.e., $\chi_\mathrm{bulk}$ $<$ 0. The fitted $\chi_\mathrm{bulk}$($T$) is roughly consistent with the total susceptibility measured at high temperatures, but doesn't exhibit a profound Curie-like tail at low temperatures (see Figure~\ref{fig3}(a)). 


No obvious magnetic hysteresis is observed, thus precluding ferromagnetic freezing down to 1.8 K (Figure~\ref{fig3}(b)). The diamagnetism is confirmed by $M$ measured above $\sim$ 10 K, as d$M$/d$H$ $<$ 0.  Below $\sim$ 10 K, the paramagnetism of tiny impurities (i.e. d$M$/d$H$ $>$ 0) dominates at $H$ $<$ $H_\mathrm{p}$, whereas the bulk diamagnetism dominates at $H$ $>$ $H_\mathrm{p}$ after the polarization of the impurity spins.  At $\mu_0H_\mathrm{p}$ = 2-3 T and 1.8 $\le$ $T$ $\le$ 10 K, $M$ exhibits a broad hump with the maximum magnetization, $M_0$ $\le$ 0.002$\mu_\mathrm{B}$/Ni, suggesting that most of Ni atoms are in the nonmagnetic state in our samples of Zr$_2$NiP$_2$. 

\begin{figure}[t]
\begin{center}
\includegraphics[width=9.6cm,angle=0]{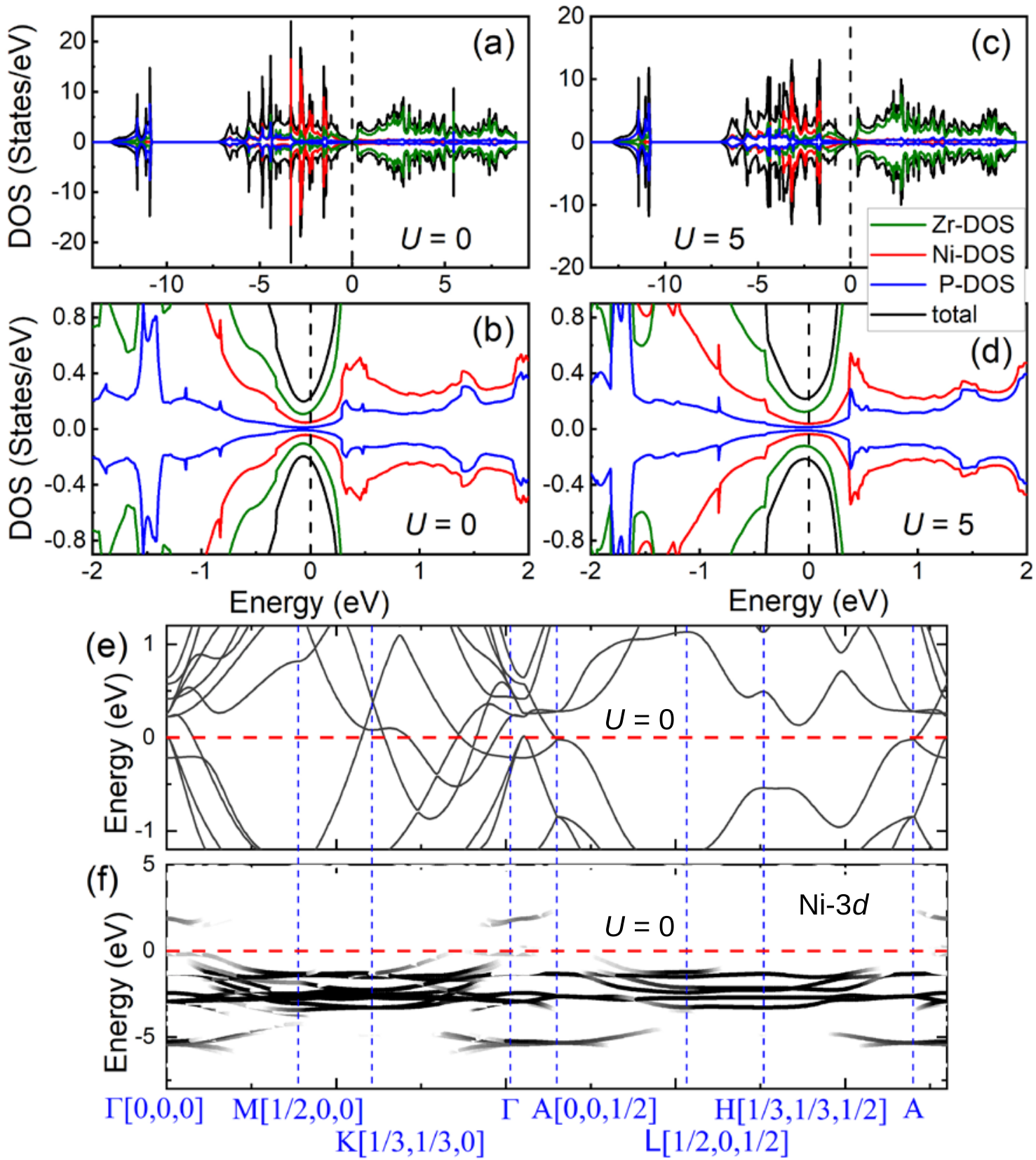}
\caption{
The total, Zr-, Ni-, and P-DOS of Zr$_2$NiP$_2$ calculated by DFT with $U$ = (a)  0 eV and (c) 5 eV, respectively. (b,d) Partial enlarged views of DOS near the Fermi energies ($E_\mathrm{F}$ = 0 eV). The spin-up and spin-down states are distinguished by positive and negative DOS, respectively, for clarity. (e) The band structure. (f) The projection to the Ni-3$d$ states.}
\label{fig4}
\end{center}
\end{figure}

\subsection{First-principles calculations}

To understand the observed metallic and nonmagnetic characters, we calculated the density of states (DOS) by DFT+$U$ for Zr$_2$NiP$_2$. Figure~\ref{fig4} shows the total and partial DOS of Zr$_2$NiP$_2$ calculated with $U$ = 0 and 5 eV, respectively. Several dispersive bands passing over the Fermi energy ($E_\mathrm{F}$ = 0 eV) are clearly observed with significant DOS (see Figure~\ref{fig4}(b), (d), and (e)), suggesting a typically metallic character of Zr$_2$NiP$_2$. The DOS at $E$ $\ge$ $E_\mathrm{F}$ is dominantly contributed by Zr atoms, which accounts for the metallic nature.

The core diamagnetic contribution to the magnetic susceptibility can be estimated as $\chi_\mathrm{D}$ $\sim$ $-$MW$\times$2$\pi\times$10$^{-6}$ cm$^3$/mol (SI unit)~\cite{kahn1993molecular} $\sim$ $-$2$\times$10$^{-3}$ cm$^3$/mol-Ni that is close to the experimental value $\chi_\mathrm{bulk}$ $\sim$ $-$3$\times$10$^{-3}$ cm$^3$/mol-Ni (see Figure~\ref{fig3}(a)), where MW = 303 is the molecular weight of Zr$_2$NiP$_2$. Therefore, it is reasonable to estimate that the Pauli susceptibility ($\chi_\mathrm{Pauli}$) should be less than $\sim$ 3$\times$10$^{-3}$ cm$^3$/mol-Ni. We can obtain the DOS at the Fermi level $D$($E_\mathrm{F}$) = $\chi_\mathrm{Pauli}$/($N_\mathrm{A}\mu_0\mu_\mathrm{B}^2$) $<$ 7 States/eV, which does not contradict with our DFT results. Alternately, the Pauli susceptibility can be estimated as $\chi_\mathrm{Pauli}$ =  $N_\mathrm{A}\mu_0\mu_\mathrm{B}^2$$D$($E_\mathrm{F}$) = 2$\times$10$^{-4}$ cm$^3$/mol-Ni,  where $D$($E_\mathrm{F}$) $\sim$  0.5 States/eV is calculated by DFT (see Figure~\ref{fig4}(b) and (d)). Usually, it is very difficult to precisely measure such a small Pauli susceptibility.

In the DFT+$U$ simulations, we considered both the initial antiferromagnetic and ferromagnetic structures for Ni atoms, $|\uparrow\downarrow\rangle$ and $|\uparrow\uparrow\rangle$ with a large magnetic moment $m$ = 2$\mu_\mathrm{B}$, but finally got the same nonmagnetic state with $m$ $<$ 0.01$\mu_\mathrm{B}$ after optimizations. Moreover, the spin-up and spin-down DOS are almost symmetric (see Figure~\ref{fig4}), which naturally explains the non-magnetism observed in Zr$_2$NiP$_2$ (see above)~\cite{Ali2020,Cheng2013}. The number of states (occupancy) integrated the Ni DOS (Figure~\ref{fig4}(a)) over $-$15 $\le$ $E$ $\le$ 0 eV (Fermi energy $E_\mathrm{F}$ = 0 eV) is calculated to be 9.5, and thus the Ni valence may be $\sim$ +0.5 in Zr$_2$NiP$_2$. The band structure projected to the Ni-3$d$ states is shown in Figure~\ref{fig4}(f), most of the weight is below the Fermi level, suggesting Ni tends to have the nonmagnetic electronic configuration $d^{10}$.

There exists no essential difference between the DFT results calculated with $U$ = 0 and 5 eV. Furthermore, we also performed the exactly similar DFT+$U$ simulations for the other members of the family,  Ln$_2$NiX$_2$ with Ln = Y, In, Lu, Zr and X = P, As, and obtained similar electrical and magnetic properties, i.e., metallic nature and nonmagnetic state of Ni. These results imply that the metallic and nonmagnetic ground state of Zr$_2$NiP$_2$ is fairly stable.

\section{Conclusions}

Polycrystalline and nearly pure-phase Zr$_2$NiP$_2$ has been successfully synthesized by the solid-state method. Structural characterization shows that the Ni atoms are arranged on a geometrically perfect triangular lattice and spatially separated by the double layers of ZrP$_6$ octahedra. The first electrical transport and magnetic property measurements consistently demonstrate that the Ni atoms are in the nonmagnetic state. The main actors of metallicity of Zr atoms are responsible for the observed metallic behavior, as revealed by the first-principles calculations. Our work should pave the way to exploring the intriguing properties  of the Ni-based triangular-lattice family Ln$_2$NiX$_2$, as well as other quasi-two-dimensional materials.   

\section*{Acknowledgement}

We thank Prof. Kan Zhao, Ying Li, and Bin Xi for helpful discussion. This work was supported by the Fundamental Research Funds for the Central Universities, HUST: 2020kfyXJJS054.

\end{document}